\documentclass{article}
\usepackage{setspace}
\usepackage[utf8]{inputenc}
\usepackage{amsmath}
\usepackage{slashed}
\usepackage{amssymb}
\usepackage{bbold}
\usepackage{halloweenmath}
\usepackage{graphicx}
\usepackage{authblk}
\usepackage{caption}
\graphicspath{ {./images/} }
\usepackage{comment}
\usepackage{tikz}
\usepackage{geometry}
\usepackage{hyperref}
\hypersetup{ 
colorlinks=true,
linkcolor=blue,
filecolor=magenta,
urlcolor=blue,
citecolor=blue,
pdfpagemode=Fullscreen
}

\usepackage[normalem]{ulem}
\usepackage{enumerate}
\usepackage[shortlabels]{enumitem}

\title{An Analytic Yang-Mills Vacuum Calculation in \(3+1d\)}
\author[1]{Seth Grable}

\affil[1]{Department of Physics, University of Colorado Boulder, Colorado 80309, USA}
\date{July, 2024}

\begin{document}
\begin{spacing}{1.5}
\maketitle

\begin{abstract}
    I present an analytic framework for \(SU(N)\) Yang-Mills theory in the four-dimensional continuum. Background and effective field theory techniques are used to include non-perturbative contributions from cubic and quartic interactions. This approach is inspired by Savvidy, who claims first-order contributions from quartic interactions stabilize IR divergence found at one-loop order, making IR finite Yang-Mills calculations possible. I assess the validity of this claim and discuss the implications of my findings.
\end{abstract}

\section{Analytic difficulties of Yang-Mills Theories}
\paragraph{}

Asymptotic freedom of non-abelian gauge theories explains phenomena like confinement and Bjorken scaling, providing an analytic foundation for Quantum Chromodynamics (QCD) \cite{coleman1973price,gross1973ultraviolet}. However,  perturbation theory fails to capture non-perturbative effects such as resummed diagram classes, which generate self-energy contributions, and becomes invalid at confinement energy scales due to negative beta functions.

Nonetheless, lattice QCD has successfully calculated glueball masses, hadronic spectrum, and the critical temperature in QCD \cite{fodor2012light,laermann2003lattice}. Yet, a disconnect remains between lattice results and analytic methods. This, along with a lack of computational understanding of finite-density QCD due to the sign problem, makes non-perturbative Yang-Mills calculations highly desirable.

Non-perturbative field theories date back to classical background field configurations of QED in Euler-Heisenberg Lagrangians. Schwinger then introduced the proper-time formalism, giving way to a large class of possible effective theories
\cite{heisenberg2006consequences,dunne2005heisenberg,schwinger1951gauge}. The background field method was further developed by DeWitt as a theory of quantum gravity \cite{dewitt1967quantum} and then applied to non-abelian gauge theories by t'Hooft \cite{t1973algorithm}. This method greatly simplifies beta-function analysis of non-abelian theories and has been shown to match perturbation theory at two-loop order by Jack, Osborn, and Abbott \cite{jack1982two,abbott1981background}, and was employed by Gross and Dashen to perform lattice calculations for general \(SU(N)\) models \cite{dashen1981relationship,abbott1981background}. Moreover, many authors have studied Yang-Mills theories with covariantly constant background field strength tensors \cite{savvidy1977infrared,leutwyler1981constant,nielsen1978unstable,yildiz1980vacuum,t1976computation}, revealing vacuum instabilities arising from zero modes in the spectrum of one-loop effective theories.
In continuing this work \cite{grable2023elements} proved that infrared instabilities cancel between gauge and matter contributions in massless twelve-flavor QCD, and more generally, Savvidy \cite{savvidy2023stability} was able to mitigate instabilities using select quartic interaction contributions.

In this article, I calculate an effective Yang-Mills Lagrangian by incorporating linear first-order contributions from cubic and quartic interactions. A modified Hubbard-Stratonovich transformation, similar to those found in large-\(N\) calculations \cite{Romatschke2022,grable2022theremal,romatschke2024mass} introduces an auxiliary field that captures these interactions at first order. Solutions to gap equations for the auxiliary field and the background gauge field collectively provide first-order contributions to the gluon self-energy in a pure gauge theory.  

Section \ref{Sec 3} presents an outline of the background field method. Section \ref{Sec 4} gives an eigenspectrum analysis of operators found in an effective Yang-Mills Lagrangian at quadratic order. Section \ref{Sec 5} uses a Hubbard-Strotanovitch transformation to set up an effective Lagrangian that includes cubic and quartic interactions at first order. Section \ref{Sec 6} presents detailed calculations of the effective theory and its associated gap equations, and compares these calculations with existing lattice data.

\section{The Background Field Set up for Yang Mills}\label{Sec 3}
\paragraph{}
Consider the Yang-Mills Lagrangian density for general SU(N) given by
\begin{equation}\label{eq 1}
    L=\frac{1}{4}(\mathcal{F}_{\mu\nu}^a)^2.
\end{equation}
 The field strength tensor and covariant derivative are defined in the adjoint representation as:
\begin{equation}\label{eq 2}
\begin{split}
   & \mathcal{F}_{\mu\nu}^a = \partial_\mu\delta^{ac} A^c_\nu(x) -\partial_\nu\delta^{ac} A^c_\mu(x) + g_0f^{abc}A^b_\mu(x)A^c_\nu(x)\\&
   D^{ac}_\mu = \partial_\mu\delta^{ac} + g_0f^{abc}A^b_\mu(x),
\end{split}    
\end{equation}
where \(f^{abc}\) are the structure constants of some SU(N) algebra, and \(g_0\) is the bare Yang-Mills coupling constant. In Euclidean space with the temporal direction compactified on the thermal cylinder, the partition function is \footnote{The measure is defined as \( \mathcal{D}A\equiv \prod_{x}\prod_{a}\prod_{\mu} d A^a_{\mu}(x)\). This shorthand notation is used throughout this article for single integrals and path integrals. That being, a product over vector or tensor indices is implied in the integral measure, and such indices are dropped.} \cite{laine2016basics}
\begin{equation}\label{eq 3}
    Z=\int\mathcal{D}A e^{-\frac{1}{4} \int_x (\mathcal{F}_{\mu\nu}^a)^2}.
\end{equation}
Equation \eqref{eq 3} demands that gauge fields \(A^a_\mu(x)\) are periodic in the temporal direction giving \(A^a_\mu(0,\vec{x}) = A^a_\mu(\beta,\vec{x})\) \cite{leutwyler1981constant}. Next the gauge fields are separated into a background-field \(B^a_\mu(x)\) and fluctuations \(a^a_{\mu}(x)\) such that \(A^a_\mu(x)=\frac{B^a_\mu(x)}{g_0}+a^a_\mu(x)\) where \(B^a_\mu(x)\) act as the zero modes of the fields in momentum space. With this, integrals over linear contributions of the fluctuations vanish\footnote{ \(\int_x a^a_\mu(x) = \int_x\int_k (\Tilde{A}^a_\mu(k)-B^a_\mu(k)) e^{ikx} =0\)} as do terms that go like \(B^a_\mu(k)a^b_\mu(k)\) due to orthogonality. This yields the field strength tensor as \cite{peskin2018introduction,weinberg1995quantum}:
\begin{equation}\label{eq F}
\begin{split}
   & \mathcal{F}_{\mu\nu}^a = \frac{1}{g_0}F_{\mu\nu}^a + D^{ac}_\mu\ a^c_\nu(x) -D^{ac}_\nu\  a^c_\mu(x) + g_0f^{abc}a^b_\mu(x)a^c_\nu(x)\\&
   D^{ac}_\mu = \partial_\mu\delta^{ac} + f^{abc}B^b_\mu(x),
\end{split}    
\end{equation}
where \(D^{ab}_{\mu}\) and  \(F^a_{\mu\nu}\) are functions of \(B^c_\mu(x)\). Choosing a covariantly constant self-dual field strength tensor \cite{leutwyler1981constant},
\begin{equation}\label{eq 7}
    F^a_{\mu\nu}=\begin{pmatrix}
        0&B^a&0&0\\-B^a&0&0&0\\0&0&0&B^a\\0&0&-B^a&0
    \end{pmatrix},
\end{equation} 
and a background-field configuration
 \begin{equation}\label{eq 6}
     B^a_\mu(x) = -\frac{1}{2}F^a_{\mu\nu}x_\nu,
 \end{equation}
satisfies the classical source-free Yang-Mills equations of \cite{savvidy1977infrared,savvidy2023stability}
 \begin{equation}
     D^{ab}_{\mu} F^b_{\mu\nu} =0.
 \end{equation} 
 With the gauge fixing condition of \(D^{ac}_\mu a^c_\mu(x) =0 \) \cite{t1976computation,savvidy1977infrared}, the gauge fixed Lagrangian is
\begin{equation}\label{eq a}
\begin{split}
   & \mathcal{L} = \frac{1}{4}\Big[\Big(\frac{1}{g_0}F^a_{\mu\nu} + D^{ac}_{\mu}a^c_\nu - D^{ac}_{\nu}a^c_\mu + g_0f^{abc}a_\mu^b a_\nu^c \Big)^2  + 2(D^{ac}_\mu a^c_\mu)^2 \Big] +\Bar{c}^a[(D^2)^{ac}+f^{abc}D^{gb}_\mu a_\mu^g]c^c,
\end{split}   
\end{equation}
where \(c^a\) and \(\bar{c}^a\) are Grassmann fields. Equation \eqref{eq a} is invariant to local gauge transformations \cite{peskin2018introduction}
\begin{equation}\label{transforms}
\begin{split}
        &a^a_\mu(x) \rightarrow a^a_\mu(x)- f^{abc}\beta^b(x) a^c_\mu(x)\\&
        B^a_\mu(x) \rightarrow B^a_\mu(x) + B^a_\mu(x)D_\mu \beta^a(x) \\&
        c^a(x)\rightarrow c^a(x) -f^{abc}\beta^b(x) c^c(x).
\end{split}
\end{equation}
Although generic manipulations of terms containing \(a^a_\mu\) destroys local gauge invariance with respect to \(a^a_\mu\), it is possible for the theory to remain invariant under local transformations of the background field \cite{abbott1981background}. This is perhaps advantageous to gauge fixing in perturbation theory, which reduces the theory from being locally invariant to only globally invariant with respect to BRST symmetry.
The choice of background field \eqref{eq 6} breaks global rotational invariance. However, Savvidy \cite{savvidy2024large} has recently shown a continuous class of functions satisfying \eqref{eq 7}. Averaging over this class of solutions restores rotational invariance \cite{nielsen1979quantum,ambjorn1980formation}, and it is postulated that any solution to \eqref{eq 7} in the form of a non-trivial covariantly constant background configuration generates a universal effective action \cite{savvidy2024large}. This is perhaps, likewise advantageous to explicit Lorentz symmetry breaking, originally pointed out by Wilson \cite{wilson1974confinement}, that occurs with the implementation of a finite cutoff scale in lattice calculations.
Considering kinetic terms in the action such as \((D^{ac}_{\mu}a^c_\nu)^2\), integration by parts can be performed over the gauge fluctuations \(a^a_\nu\) as they now transform as matter fields in the adjoint representation \cite{peskin2018introduction}. Equation \eqref{eq 3} augmented with ghost fields \(c^a(x)\) now reads 
\begin{equation}\label{eq 12}
    Z=\int B\int \mathcal{D}a\mathcal{D}\bar{c}\mathcal{D}c e^{-\int_x \frac{1}{4g^2_0}(F^a_{\mu\nu})^2-S_0-S_I}
\end{equation}
where \cite{grable2023elements} \cite{savvidy2023stability}
\begin{equation}\label{}
    \begin{split}\label{eq 13}
        &S_0 = \int_x \frac{1}{2}a^a_\mu\Big[-(D^2)^{ac}\delta_{\mu\nu} - 2F_{\mu\nu}^bf^{abc}\Big]a^c_\nu + \Bar{c}^a(-D^2)^{ac}c^c\\&\&\\&
       S_I=\int_x g_0(D_\mu a_\nu^a)f^{abc} a_\mu^b a_\nu^c -\Bar{c}^a (f^{abc}D^{gb}_\mu) a_\mu^g c^c+ \frac{g_0^2}{4}(f^{abc}a_\mu^b a_\nu^c)^2. 
    \end{split}
\end{equation}
The quadratic terms are relabeled as \(\theta_{\text{Glue}}\equiv-(D^2)^{ac}\delta_{\mu\nu} - 2F_{\mu\nu}^bf^{abc}\), and \(\theta_{\text{Ghost}}\equiv-(D^2)^{ac} \) presenting the one-loop effective theory as \cite{peskin2018introduction}
\begin{equation}\label{one loop}
  Z\approx \int dB e^{-\frac{\beta V}{4g^2_0}(F^a_{\mu\nu})^2-\frac{1}{2}\ln\det(\theta_{\text{Glue}})+\ln\det(\theta_{\text{Ghost}})}.
\end{equation}
Although gauge invariance with respect to the fluctuations is lost, the one-loop theory remains gauge invariant with respect to local transformations of the background ground field as all functions of \(B^a_\mu(x)\) in \ref{one loop} are multiples of \(D^{ac}_\mu\) and \(F^a_{\mu\nu}\) \cite{abbott1981background}. However, as noted by Savvidy, Leutwyler, and others \cite{savvidy1977infrared,leutwyler1981constant,yildiz1980vacuum}, dropping gauge fluctuations beyond quadratic order generates an IR divergence in \(\ln\det(\theta_{\text{Glue}})\), indicating the need for regulating terms from higher-order interactions.

\section{The One-Loop Eigenspectrum}\label{Sec 4}
\paragraph{}
To find the eigenspectrum of \(\theta_{\text{Glue}}\) the covariant derivative and the field strength tensor are considered separately. Because \(-(D^2)^{ac}\delta_{\mu\nu}\) and \(2F_{\mu\nu}^bf^{abc}\) are simultaneously diagonalizable in Lorentz and color space \cite{yildiz1980vacuum,grable2023elements} the sum of the eigenvalues of \(-(D^2)^{ac}\delta_{\mu\nu}\) and \(2F_{\mu\nu}^bf^{abc}\) equals the eigenvalues of \(-(D^2)^{ac}\delta_{\mu\nu}+2F_{\mu\nu}^bf^{abc}\). Equations \eqref{eq F} and \eqref{eq 6} give the covariant derivative in terms of the field strength tensor as
\begin{equation}
    D^{ac}_\mu  = \partial_\mu \delta^{ac} +\frac{i}{2}A^{ac}F_{\mu\nu}x_\nu
\end{equation}
where \(\textbf{A}\) is the hermitian matrix consisting of the sum generators in the adjoint representation with components \(A^{ac}\). In letting \(B^bf^{abc}= BA^{ac}\) I am making a choice of background field such that \(B^a=B^c \ \forall a,c\). Letting the Lorentz index run from 0 to 3 gives \cite{savvidy2023stability}
\begin{equation}\label{eq 17}
\begin{split}
        -(D^2)^{ac} =  &-(\partial^2_0)\delta^{ac}  -(\partial^2_1)\delta^{ac}  + i(AB)^{ac}\big(x_1\partial_0-x_0\partial_1\big) + \frac{1}{4}(A^2B^2)^{ac}\big(x_0^2+x_1^2 \big)\\&
    -(\partial^2_2)\delta^{ac}  -(\partial^2_3)\delta^{ac}  + i(AB)^{ac}\big(x_3\partial_2-x_2\partial_3\big) + \frac{1}{4}(A^2B^2)^{ac}\big(x_2^2+x_3^2\big).
\end{split}       
\end{equation}
 Now consider the operators
\begin{equation}\label{eq 19}
\begin{split}
    & c_\mu =\Bigg[ \partial_\mu\ \mathbb{1} + \frac{1}{2}(B\textbf{A})\hat{x}_\mu\Bigg], \quad  c^{\dagger}_\mu = \Bigg[-\partial_\mu\ \mathbb{1} + \frac{1}{2}(B\textbf{A}) \hat{x}_\mu\Bigg],
\end{split}
\end{equation}
where \( \mathbb{1}\) is the identity in color space. Using the canonical commutation relation \([\hat{x}_\mu, -i\partial_\nu]=i\delta_{\mu\nu}\) it can be shown that \([c_\mu,c^{\dagger}_\mu]=B\textbf{A}\), giving \cite{leutwyler1981constant}
\begin{equation}
\begin{split}\label{eq opps}
     -(D^2) =&\Big[\big(c^{\dagger}_0+ic^{\dagger}_1\big)\big(c_0-ic_1\big) + B\textbf{A}\Big] +  \Big[\big(c^{\dagger}_2+ic^{\dagger}_3\big)\big(c_2-ic_3\big) + B\textbf{A}\Big].
\end{split}
\end{equation}
 The commutation relation of \((c^{\dagger}_0+ic^{\dagger}_1\big)\) and \(\big(c_0-ic_1\big)\) is
\begin{equation}\label{eq com}
    [\big(c_0-ic_1\big),(c^{\dagger}_0+ic^{\dagger}_1\big)] =2B\textbf{A}.
\end{equation}
Thus, \((c^{\dagger}_0+ic^{\dagger}_1\big)\big(c_0-ic_1\big)\) effectively acts as a double number operator, and \(-(D^2)\) has the form of harmonic oscillators in the \(0-1\) and \(2-3\) planes. The eigenspectrum \(\Omega\) of \(-(D^2)\) is  
\begin{equation}\label{eq 23}
    \Omega= \sum_{m,n} B\textbf{A}(2n+1) + B\textbf{A}(2m+1).
\end{equation}
 Finally, the eigenvalues of \(F_{\mu\nu}\) come in complex pairs of \(\pm i B\) giving the total eigenspectrum of \(\theta_{\text{Glue}}\) as \cite{savvidy2023stability},
\begin{equation}\label{eq 20}
\begin{split}
    &(\Lambda^a_+)_{m,n} = (2n+1)B\lambda^a +(2m+1)B\lambda^a +2B\lambda^a \\&
    (\Lambda^a_-)_{m,n} = (2n+1)B\lambda^a+(2m+1)B\lambda^a-2B\lambda^a,
\end{split}
\end{equation}
with \(m,n\in \mathbb{N}\) and \(\lambda^a\) the eigenvalues of \(\textbf{A}\). The eigenstates of \(-(D^2)\) are given as
\begin{equation}\label{eq states}
    b^a_{\mu}(x)_{mn} = \big((c^{\dagger}_0)^{ab}+i(c^{\dagger}_1)^{ab}\big)^n\big((c^{\dagger}_2)^{ab}+i(c^{\dagger}_3)^{ab}\big)^m b^a_{\mu}(x)_{00},
\end{equation}
where the groundstate \(b_{00}\) is given by
\begin{equation}\label{eq ground}
   b^a_{\mu}(x)_{00} = (e^{-\frac{1}{4} B\lambda^a(x_\nu)^2})_{\mu}.
\end{equation}
The non-trivial spectrum of  \(\theta_{\text{Glue}}\) contains a zero eigenvalue  where \(m=n=0\), explicitly generating an IR divergence in the one-loop approximation \cite{savvidy2023stability,nielsen1978unstable,leutwyler1981constant}. As shown by Savvidy \cite{savvidy2023stability}, contributions from cubic and quartic interactions regulate this divergence. In the following section \ref{Sec 5} contributions from cubic and quartic interactions are added using an auxiliary field \(\Delta\). As will be seen in section \ref{Sec 6}, with the use of \(B\) and \(\Delta\), an effective form of \eqref{eq 13} can be calculated with a saddle point approximation that is asymptotically dominant in the large \(\beta V\) limit.

\section{Vacuum Calculations with the Use of an Auxiliary Field}\label{Sec 5}
\paragraph{}
Starting with the gauge fixed partition function and dropping the linear coupling in \(a^a_\mu\) to the ghost fields gives 
\begin{equation}\label{eq 8}
    Z_0= \int dB \mathcal{D}a \mathcal{D}c\mathcal{D}\Bar{c} e^{-\int_x (\frac{1}{4g^2_0}F^a_{\mu\nu})^2 -S_0 - S_I}
\end{equation}
where
\begin{equation}\label{eq int}
    \begin{split}
       & S_0=\int_x \frac{1}{2} a^a_\mu[\theta_{\text{Glue}}]a^c_\nu + \Bar{c}^a[\theta_{\text{Ghost}}]c^c\\&
       S_I=\int_x -g_0(D_\mu a_\nu^a)f^{abc} a_\mu^b a_\nu^c + \frac{g^2_0}{4}(f^{abc}a_\mu^b a_\nu^c)^2.
    \end{split}
\end{equation}
To integrate the interaction terms at \(R_0\) level \cite{romatschke2019simple} a Hubbard Stratonovich transformation is applied to \eqref{eq 8} of the form \cite{grable2022theremal,romatschke2019finite},
\begin{equation}\label{hub}
    \int^{\infty}_{-\infty}\mathcal{D}\sigma\int^\infty_0 \mathcal{D\xi} \text{Re}[e^{i\int_x \xi^{a}_{\mu\nu}(\sigma^{a}_{\mu\nu} - f^{abc}a^b_\mu a^c_\nu)}] = 1,
\end{equation}
 under the ansatz that \(\xi^{a}_{\mu\nu}\) is diagonal in Lorentz space, the transformation acts only on such terms, and the remaining terms in \(S_I\) not diagonal in Lorentz space, are discarded. Further, the integral measures in \eqref{hub} only contain terms where \(\mu=\nu\) as does \(\sigma^a_{\mu\nu}\). The effective action of the gluon contribution is now
\begin{equation}
  S^{\text{glue}}_{\text{eff}}=  -\int_x \frac{1}{2}a^a_\mu[\theta_{\text{Glue}}^{ac}-i\xi^{b}_{\mu\nu}f^{abc}]a^c_\nu  - g_0(D_\mu a_\nu^a)\sigma^{a}_{\mu\nu} + (\frac{g_0}{2}\sigma^{a}_{\mu\nu})^2 - i\xi^{a}_{\mu\nu}\sigma^{a}_{\mu\nu}.
\end{equation}
The term \(- g_0(D_\mu a_\nu^a)\sigma^{a}_{\mu\nu}\) is diagonal in Lorentz space due to \(\sigma^a_{\mu\nu}\) and therefore vanishes under the gauge fixing condition. Thus the cubic interaction terms do not contribute to the scalar gluon self-energy at first order. Savvidy similarly claims the cubic interactions do not contribute to regulating the zero mode divergence at first order \cite{savvidy2023stability}. Integrating out \(\sigma^a_{\mu\nu}\),  letting \(\xi^a_{\mu\nu}=\bar{\xi}^a_{\mu\nu}+(\xi')^a_{\mu\nu}(x)\), and discarding fluctuations \((\xi')^a_{\mu\nu}(x)\) results in
\begin{equation}\label{eq 36}
  S^{\text{glue}}_{\text{eff}}=  -\int_x \frac{1}{2}a^a_\mu[-(D^2)^{ac}\delta_{\mu\nu} + 2A^{ac}F_{\mu\nu}-\Bar{\xi}_{\mu\nu} A^{ac}]a^c_\nu +\frac{(\Bar{\xi}^a_{\mu\nu})^2}{g^2_0},
\end{equation}
The variable substitution \(\Bar{\xi}^a_{\mu\nu}\rightarrow B^a\Delta\delta_{\mu\nu}\) gives the effective action of
\begin{equation}\label{eq 36}
  S^{\text{Glue}}_{\text{eff}}=  -\int_x \frac{1}{2}a^a_\mu[-(D^2)^{ac}\delta_{\mu\nu} + 2A ^{ac}F_{\mu\nu}+\Delta B A^{ac}
 \delta_{\mu\nu}]a^c_\nu +\frac{(B^a\Delta\delta_{\mu\nu})^2}{g^2_0}.  
\end{equation}
Taking the saddle point approximation for \(B\) and \(\Delta\) the partition function is now.
\begin{equation}\label{eq Z}
    Z= \int\mathcal{D}a\mathcal{D}c\mathcal{D}\Bar{c}\text{ Re}\Big[ e^{-\int_x \frac{1}{4g^2_0}(\Bar{F}^{a}_{\mu\nu})^2+\frac{(B^a\Delta\delta_{\mu\nu})^2}{g^2_0}+a^a_\mu\big[-(D^2)^{ac}\delta_{\mu\nu} +2A^{ac}F_{\mu\nu} + \Delta B A^{ac}
 \delta_{\mu\nu}\big]a^c_\nu +\bar{c}^a\big[(-(D^2)^{ac}\big]c^c}\Big]_{\Bar{B},\Bar{\Delta}}.
\end{equation}
 where \(\Bar{\Delta}\) and \(\Bar{B}\) are saddle points of their respective functions in \eqref{eq Z}. The form of \eqref{hub} constrains the saddle equations of \(\Delta\) to the positive real axis, likewise constraining the eigenspectrum of \(\theta^{R_0}_\text{Glue}\) to be semi-positive definite. This must be the case as the original form of \eqref{eq 3} is semi-positive definite. The constant \(\frac{(\Bar{B}^a\bar{\Delta}\delta_{\mu\nu})^2}{g^2_0}\) can be removed with a counter term giving an effective \(Z\) as
\begin{equation}\label{eq 41}
    Z= e^{-\frac{\beta V}{g_0^2}(F^a_{\mu\nu})^2-\frac{1}{2}\ln\det\big[\theta^{R_0}_\text{Glue}\big]+\ln\det\big[\theta_\text{Ghost}\big]}\Big|_{\bar{B}\bar{\Delta}}
\end{equation}
where \(\theta^{R_0}_\text{Glue}=-(D^2)^{ac}\delta_{\mu\nu} + 2 A^{ac} F_{\mu\nu} + \Delta B A^{ac}\delta_{\mu\nu}\). Similar to the one-loop theory in \ref{one loop} the effective action in \ref{eq 41} remains gauge invariant with respect to local transformations of \(B_\mu^a(x)\) but not \(a^a_\mu(x)\) as all functions of \(B_\mu^a(x)\) in \ref{eq 41} are powers of \(D^{ac}_\mu\) and \(F^a_{\mu\nu}\) and the term \(a^a_\mu(x) \Delta B A^{ac}\delta_{\mu\nu}a^c_\nu(x)\) only transforms locally with respect to \(a^a_\mu(x)\).

\section{Calculating the Effective Action}\label{Sec 6}
\paragraph{}
To calculate \(\ln\det\theta_{\text{Ghost}}\) I will use zeta function regularization of the form \cite{bertlmann2000anomalies}
\begin{equation}
    \ln \det \theta =-\Big[\frac{d}{ds} \frac{1}{\Gamma[s]}\int_0^\infty d\tau\tau^{s-1} K_{\theta}\Big]_{s=0},
\end{equation}
where
\begin{equation}
    K_{\theta} = \text{Tr}^{ab}_{\mu\nu}\sum_{n,m}e^{-\tau\theta/\mu^2}
\end{equation}
is the heat kernel of the operator \(\theta\) and \(\mu\) is an arbitrary renormalization scale. The heat kernels are given as \cite{grable2023elements,savvidy2023stability}
\begin{equation}
\begin{split}
  &  K_{\text{Ghost}} = \sum_a\beta V \frac{(B\lambda^a)^2}{16\pi^2}\left[\frac{1}{\sinh^2(\frac{B\lambda^a\tau}{\mu^2})}\right]
  \\&
  \&\\&
  K^{R_0}_{\text{Glue}} =\sum_a\beta V \frac{(B\lambda
^a)^2}{4\pi^2}\left(e^{-\frac{\tau B 
\lambda^a \Delta}{\mu^2}}\right)\left[2+\frac{1}{\sinh^2(\frac{B\lambda^a\tau}{\mu^2})}\right].
\end{split} 
\end{equation}
The ghost field contribution gives
\begin{equation}\label{eq 30}
\begin{split}
   & \ln\det{\theta_{\text{Ghost}}} 
  = - \sum_a \beta V \frac{(B\lambda^a)^2}{48 \pi^2} \left[ \ln\left(\frac{B\lambda^a}{\mu^2}\right)+\ln\left(\frac{2e}{A^{12}}\right)\right].
\end{split}   
\end{equation}
where \(A\) is the Glaisher constant\footnote{ For the remainder of this paper the color matrix \textbf{A} does not appear, only components \(\lambda^a\) are relevant. In the following calculations any use of the symbol  \(A\) is the Glaisher constant.}. For the gluon contribution, the exponential dependence on \(\Delta\) is expanded giving
\begin{equation}\label{det glue}
\begin{split}
    &-\frac{1}{2}\ln\det[\theta^{R_0}_\text{Glue}] =\sum_a\frac{\beta V (B\lambda^a)^2 }{8 \pi ^2}\times\frac{d}{ds}\Bigg[\frac{1}{\Gamma(s)}\Bigg(2\int^\infty_0 d\tau \tau^{s-1} e^{-\tau\frac{ B\lambda^a\Delta}{\mu^2}}\\&+ \int^\infty_0 d\tau \sum^{2}_{n=0}\frac{\tau^{s+n-1}(-\frac{B\lambda^a\Delta}{\mu^2})^n}{n!} \frac{1}{\sinh^2\left(\frac{B\lambda^a\tau}{\mu^2}\right)} +\int^\infty_0 d\tau \sum^{\infty}_{n=3}\frac{\tau^{s+n-1}(-\frac{B\lambda^a\Delta}{\mu^2})^n}{n!} \frac{1}{\sinh^2\left(\frac{B\lambda^a\tau}{\mu^2}\right)}\Bigg)\Bigg]_{s=0}.
\end{split}   
\end{equation}
The first integral in \eqref{det glue} vanishes in the IR, curing the one-loop IR divergence. The second integral containing \(n<3\) contributions gives two UV divergences, one from the \(n=0\) term, and one from the \(n=2\) term. Further, the \(n=1\) and \(n=2\) terms give the only non-vanishing IR contributions. The second sum in \eqref{det glue} contains a series of UV finite contributions, and is real, and absolutely and uniformly convergent for \(\Delta \geq 0\) giving commutativity of the limits in \eqref{det glue}, collectively yielding, with the sum over color space implied,
\begin{equation}\label{eq 42}
\begin{split}
   & \frac{\ln Z}{\beta V} = -\frac{(B\lambda^a)^2}{g_0^2}-\frac{(B\lambda^a)^2}{48\pi^2}\Bigg[11 \ln \left(\frac{B \lambda^a}{\mu ^2}\right) + C +f(\Delta)+g(\Delta,B,\mu)
   \Bigg]\Bigg|_{\bar{B}\bar{\Delta}}
\end{split}    
\end{equation}
where 
\begin{equation}
\begin{split}
  & C= \ln \left(\frac{e}{2 A^{12}}\right) \\& \& \\&
 g(\Delta,B,\mu)= 3 \Delta^2\ln \left(\frac{B \lambda^a}{\mu ^2}\right)+12 \ln (\Delta )\\&
  \& \\&
  f(\Delta)= -\Delta  \left(\ln \Big(64 \pi ^6\right)-3\Delta  \ln (2)\Big)
  +12 \Delta  \ln \left(\Gamma \left[\frac{\Delta +2}{2}\right]\right)- 24 \zeta ^{(1,0)}\left(-1,\frac{\Delta +2}{2}\right).
\end{split}
\end{equation}
Evaluating \eqref{eq 42} at \(\lambda^a=0\) results in \(\frac{\ln Z}{\beta V}=0\). This leaves no gap equations for \(B\) in the \(\lambda^a=0\) color channels. However, if \(\bar{B}\rightarrow \infty\) when \(\lambda^a=0\) then \(\frac{\ln Z}{\beta V}\) goes to an undetermined constant. This indeed turns out to be the case and the gap equations for \(\bar{B}\) determine the value of \(\frac{\ln Z}{\beta V}\) when \(\bar{B}\rightarrow \infty\) and \(\lambda^a=0\).

\begin{figure}[h]
\centering
\includegraphics[width=.8\textwidth]{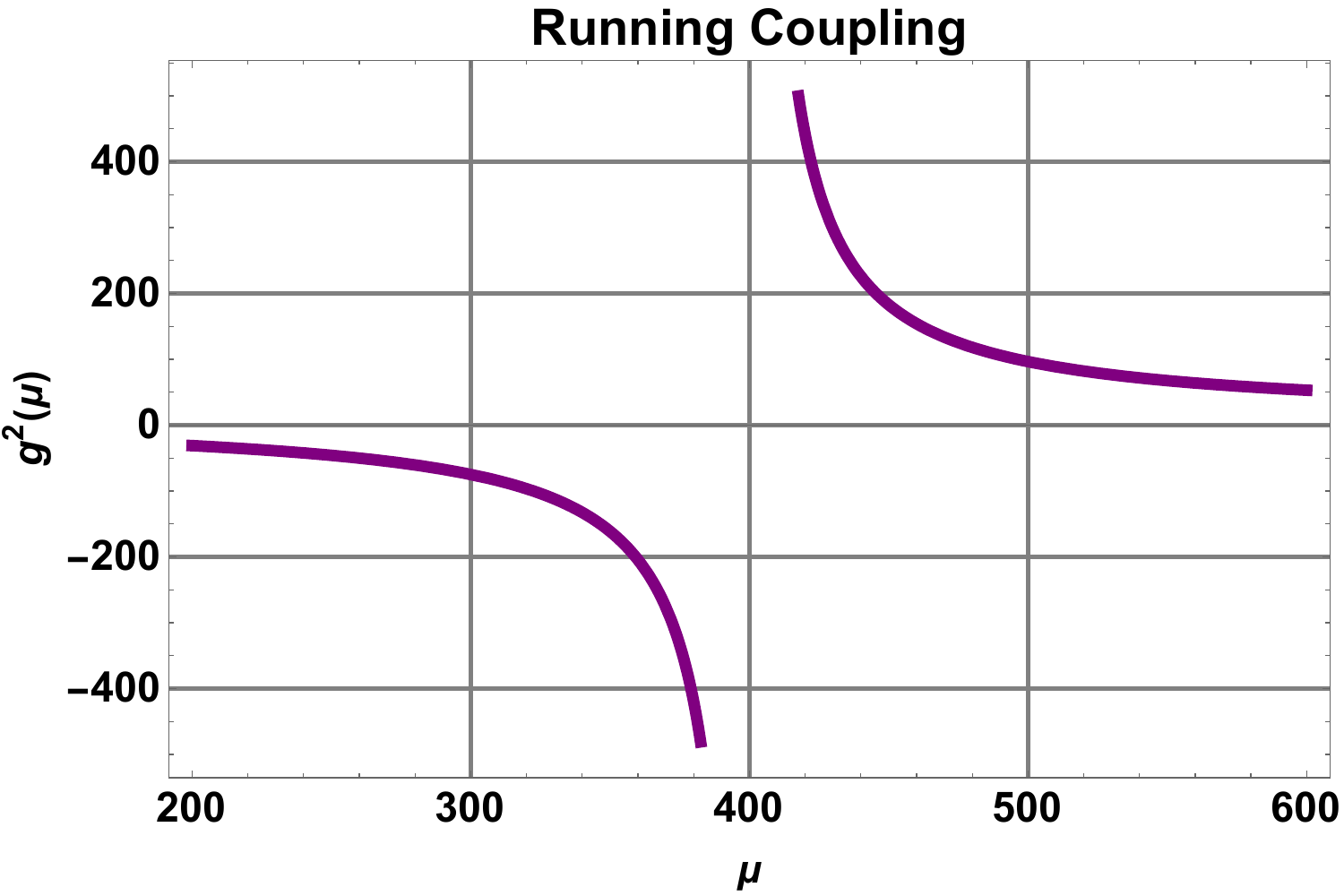}
\caption{\label{graph}The running coupling is plotted as a function of \(\mu\) with the a set Landau pole scale of \(\Lambda_\text{YM}= 400\text{MeV}\).}
\label{runnin}
\end{figure}

There are a variety of ways to regulate the divergent gluon contributions, and the self-energy of the gauge fields is numerically sensitive up to the scale of the Landau pole for each regulated Lagrangian \cite{delamotte2012introduction}. To maintain all IR physics, and remove the UV divergent, and scale-dependent physics associated with \(\Delta\) I will absorb \(g(\Delta,B,\mu)\) into the running coupling such that \cite{moshe2003quantum} 
\begin{equation}
  \frac{1}{g^2(\mu)}=  \frac{1}{g_0^2}+\frac{  1}{48 \pi ^2}g(\Delta,B,\mu),
\end{equation}
producing an effective action with all scale dependence coupled to the background field,
giving a beta-function of
\begin{equation}\label{eq 45}
    \frac{d g(\mu)}{d\ln(\mu)} =\frac{-11g^3(\mu)}{48\pi^2},
\end{equation}
which matches the perturbative beta function at first loop order times a factor of \(\frac{1}{3}\). Figure \ref{runnin} includes the full function of \eqref{eq 45}. Including the full functionality of running coupling implies the coupling does not go to infinity at the scale of the Landau pole. Instead, the limit of \(\alpha(\mu)\) as \(\mu \rightarrow \Lambda_{\text{YM}}\) does not exist, and this region contains a simple pole with a vanishing Cauchy principle value, indicating the possibility of a phase transition. A similar analysis of the complete running coupling is shown in \cite{grable2023fully, romatschke2024life, romatschke2023loophole, romatschke2023if}. With the complete function of running coupling the theory is described by a negative and weakly coupled theory deep in the IR. In this way, semi-classical expansion is well defined asymptotically to the left of the Landau pole. 

Solving \eqref{eq 45} for \(g(\mu)\) gives a renormalized pressure is of
\begin{equation}\label{eq 43}
    \begin{split}
        &\frac{\ln Z}{\beta V}=-\sum_a\frac{(B \lambda^a) ^2}{48 \pi ^2} \left[11\ln \left(\frac{B\lambda^a}{\Lambda_{\text{YM}} ^2}\right)+C +f(\Delta)\right]_{\Bar{B},\Bar{\Delta}}
    \end{split}.
\end{equation}
The function \(f(\Delta)\) can be plotted to show the existence of a clear stable minimum.
\begin{figure}[h]
\centering
\includegraphics[width=.8\textwidth]{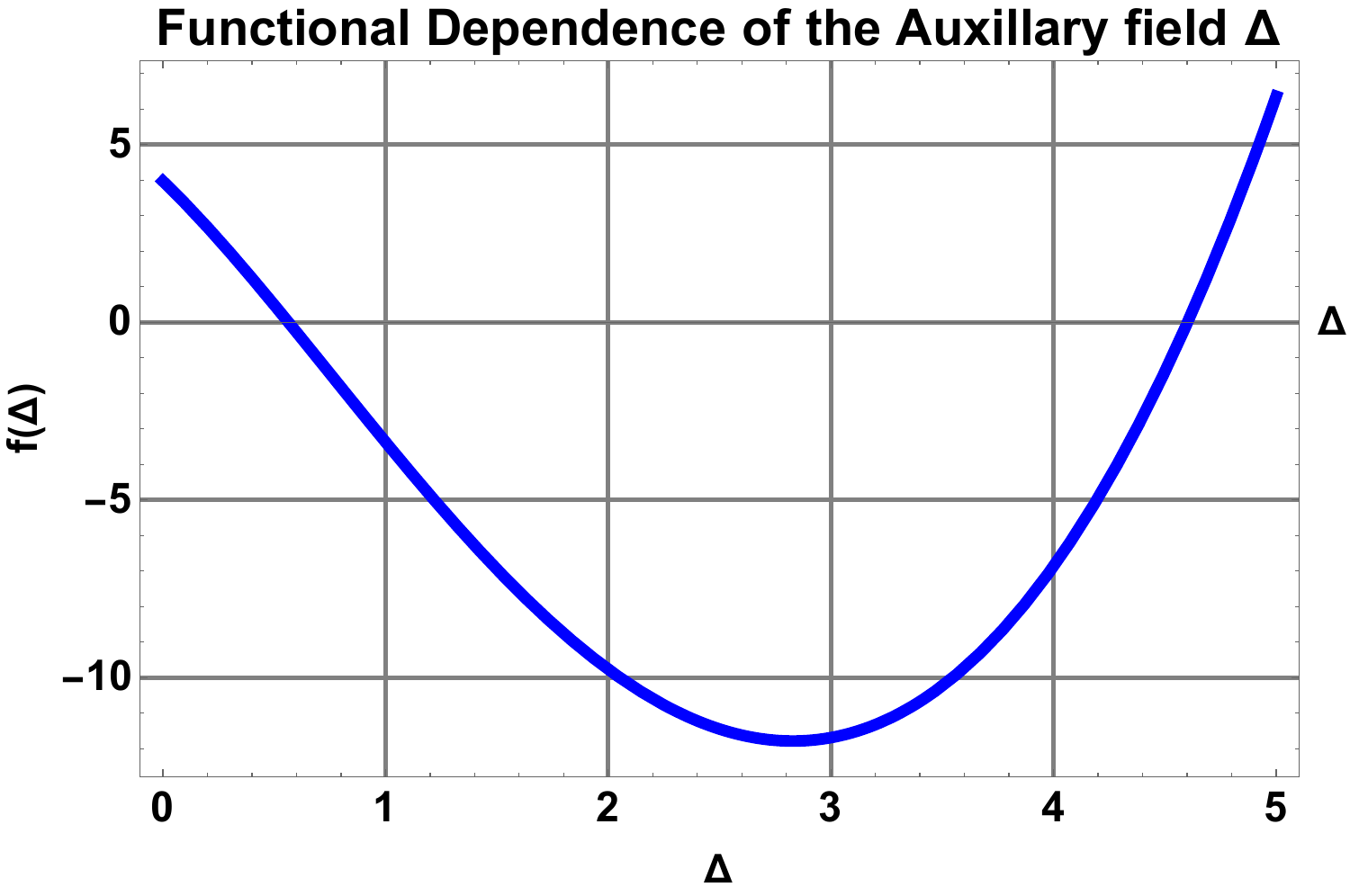}
\caption{\label{graph} The functional dependence of \(\ln Z\) on \(\Delta\) after renormalization shows a finite stable minimum.}
\label{fig: plot1}
\end{figure}
It can be numerically shown that \(f(\Delta)\) is monotonically increasing many orders of magnitude beyond the stable minimum shown in \ref{fig: plot1}.
The gap equation for \(\Delta\) is
\begin{equation}\label{eq 48}
2 \zeta ^{(1,1)}\left(-1,\frac{\Delta +2}{2}\right)+\gamma_E  \Delta -2 \log \left(\Gamma \left(\frac{\Delta +2}{2}\right)\right)-\left(\Delta  \left(H_{\frac{\Delta }{2}}+\log (2)\right)\right)+\log (2 \pi )=0,
\end{equation}
where \(H\) is the harmonic number function.
 As the gap equation for \(\Delta\) decouples from \(\lambda^a\) its saddle is identical for all color indices giving
\begin{equation}\label{gap num}
   \Bar{\Delta} \to 2.82898.
\end{equation}
Plugging \eqref{gap num} into \eqref{eq 43} gives
\begin{equation}
      \frac{\ln Z}{\beta V}=-\sum_a\frac{(B \lambda^a) ^2}{48 \pi ^2} \left[\left(14.4634\, -11 \log \left(\frac{B \lambda^a }{\Lambda_{\text{YM}}^2}\right)\right)\right].
\end{equation}
The non-zero free energy density contribution \((-\frac{\ln Z}{\beta V})^a\) is plotted, showing the existence of stable minimum, and the unstable perturbative saddle at \(\bar{B}=0\), and is likewise monotonically increasing beyond the stable minimum, structurally matching results of \cite{savvidy2020heisenberg}.
\begin{figure}[h]
\centering
\includegraphics[width=.8\textwidth]{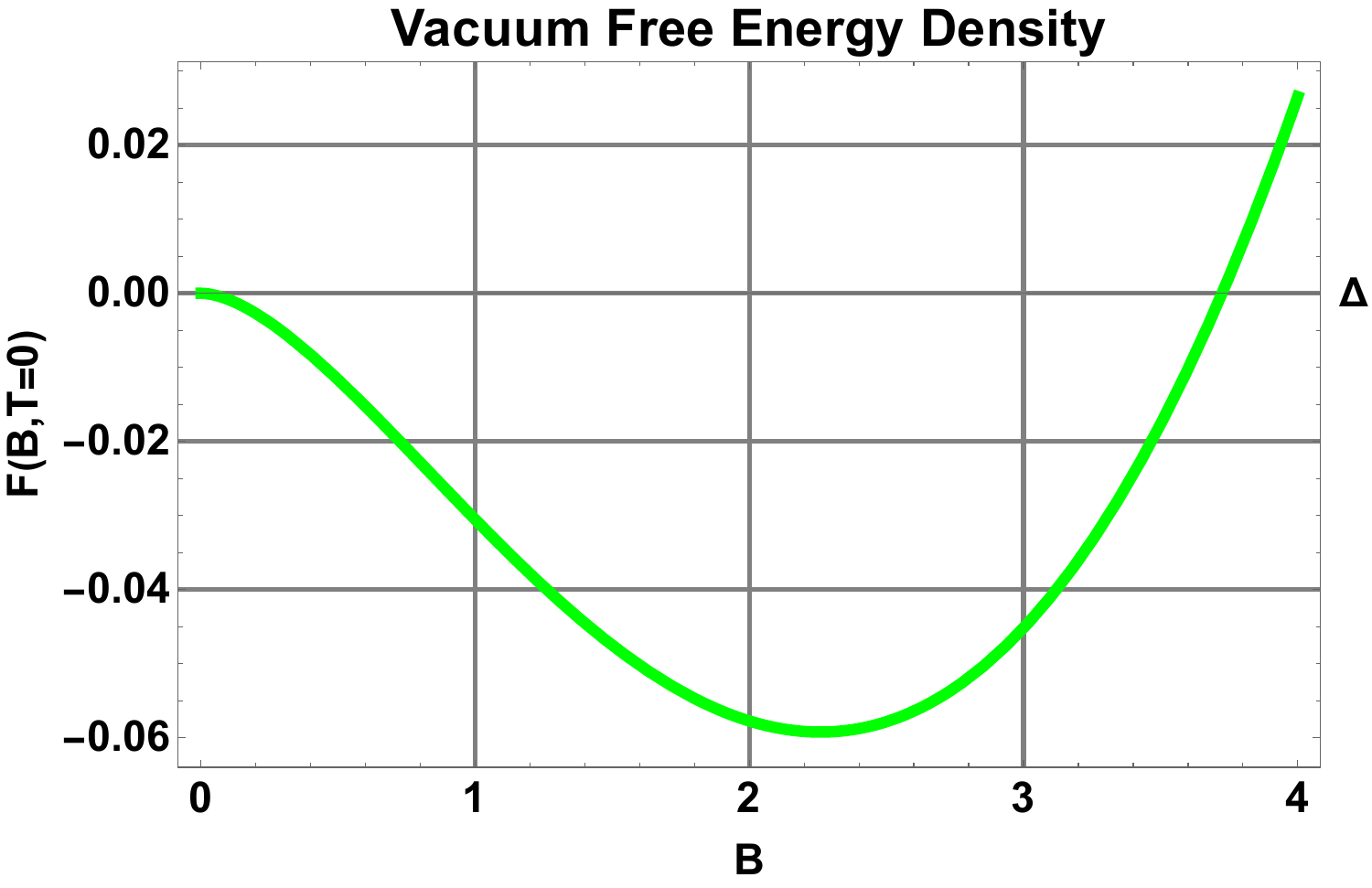}

\caption{ \label{fig: plot2} The non-zero free energy density contributions per component are shown with \(\lambda^a=1\) and \(\Lambda_{YM}=1\).}
\end{figure}
The gap equation for \(B\) is
\begin{equation}
   \left(14.4634\, -11 \log \left(\frac{B \lambda ^a}{\Lambda_{\text{YM}} ^2}\right)\right)-\frac{11 }{2}=0
\end{equation}
giving
\begin{equation}
\begin{split}
   & \Bar{B}\to \frac{2.25885 \Lambda_{\text{YM}} ^2}{\lambda^a } \text{  ,  }
   \Bar{B}\to 0,
\end{split}   
\end{equation}
where the trivial solution gives an unstable vacuum pressure, associated with the perturbative minimum. Summing over the \(SU(3)\) color index for the stable solution of \(B\) gives 
\begin{equation}
    \frac{\ln Z}{\beta V} =8\times 0.0592377 \Lambda_{\text{YM}}^4.
\end{equation} 
 Thus the final effective theory is gauge invariant with respect to the background field as the self-energy of the gauge fields is a Lorentz scalar and a gauge singlet, and the remaining terms are comprised of covariant derivatives and the field strength tensor. Now the effective mass per component of the scalar glueball can be read off of \eqref{eq 41} as
\begin{equation}\label{eq mass}
    m^a_{\text{Glue}} = \sqrt{\Bar{B}\lambda^a\Bar{\Delta}}=2.52789\Lambda_{\text{YM}}.
\end{equation}
I cannot currently make predictions on the scale of the landau pole in our theory \(\Lambda_{\text{YM}}\), nor have I done any matching condition or reparameterization analysis with other possible renormalization schemes. However, if \(\Lambda_{\text{YM}}\) fall in the range of \(400-700 \) MeV it gives a mass range of
\begin{equation}
     m^a_{\text{Glue}} \approx 1011.16\text{  to  }1769.53 \text{ MeV}
\end{equation}
agreeing with decades of quenched lattice QCD predictions for the scalar glueball mass \cite{morningstar1999glueball,ochs2013status,mathieu2009physics}.

\section{Final Considerations}
\paragraph{}

 The eigenspectrum analysis and computational methods presented in this work can be extended to include matter fields \cite{peskin2018introduction}, finite temperature, and chemical potential in QCD. One could calculate propagators of the theory in the energy-eigenbasis of \(-D^2\), and confinement parameters such as the Wilson loop. Further, Landau pole methods using the full running coupling are important as postulates from a growing literature of work on \(\mathcal{PT}\)-symmetric and negatively coupled field theories. More analysis, and experimental verification \cite{soley2023experimentally} of these methods will bolster the claims in this article. 
\section{Acknowledgments}
\paragraph{}
This work is supported by DOE award No DE-SC0017905. I am grateful for all the fruitful discussions with Paul Romatschke, Ryan Weller, Johannes Reinking, Willie Wei Su, Scott Lawrence, and Jeff Greensite.
\bibliographystyle{plain}
\bibliography{bibliography.bib}

\begin{thebibliography}{10}

\bibitem{abbott1981background}
Laurence~F Abbott.
\newblock \href{https://www.sciencedirect.com/science/article/pii/0550321381903710}{The background field method beyond one loop}.
\newblock {\em Nuclear Physics B}, 185(1):189--203, 1981.

\bibitem{ambjorn1980formation}
Jan Ambj{\o}rn and Poul Olesen.
\newblock \href{https://www.sciencedirect.com/science/article/pii/0550321380904769}{On the formation of a random color magnetic quantum liquid in QCD}.
\newblock {\em Nuclear Physics B}, 170(1):60--78, 1980.

\bibitem{bertlmann2000anomalies}
Reinhold~A Bertlmann.
\newblock {\em \href{https://books.google.com/books?hl=en&lr=&id=FC_DRRUHFXEC&oi=fnd&pg=PA1&dq=anomalies+in+quantum+field+theory&ots=PldoMsfp28&sig=1cUBC6M--cZ2myquL-NtlySIvIw#v=onepage&q=anomalies%20in%20quantum%20field%20theory&f=false}{Anomalies in quantum field theory}}, volume~91.
\newblock Oxford university press, 2000.

\bibitem{coleman1973price}
Sidney Coleman and David~J Gross.
\newblock \href{https://journals.aps.org/prl/abstract/10.1103/PhysRevLett.31.851}{Price of asymptotic freedom}.
\newblock {\em Physical Review Letters}, 31(13):851, 1973.

\bibitem{dashen1981relationship}
Roger Dashen and David~J Gross.
\newblock \href{https://journals.aps.org/prd/abstract/10.1103/PhysRevD.23.2340}{Relationship between lattice and continuum definitions of the gauge-theory coupling}.
\newblock {\em Physical Review D}, 23(10):2340, 1981.

\bibitem{delamotte2012introduction}
Bertrand Delamotte.
\newblock \href{https://link.springer.com/chapter/10.1007/978-3-642-27320-9_2}{An introduction to the nonperturbative renormalization group}.
\newblock In {\em Renormalization group and effective field theory approaches to many-body systems}, pages 49--132. Springer, 2012.

\bibitem{dewitt1967quantum}
Bryce~S DeWitt.
\newblock \href{https://journals.aps.org/pr/abstract/10.1103/PhysRev.162.1195}{Quantum theory of gravity. II. The manifestly covariant theory}.
\newblock {\em Physical Review}, 162(5):1195, 1967.

\bibitem{dunne2005heisenberg}
Gerald~V Dunne.
\newblock \href{https://www.worldscientific.com/doi/abs/10.1142/9789812775344_0014}{Heisenberg--Euler effective Lagrangians: basics and extensions}.
\newblock In {\em From Fields to Strings: Circumnavigating Theoretical Physics: Ian Kogan Memorial Collection (In 3 Volumes)}, pages 445--522. World Scientific, 2005.

\bibitem{fodor2012light}
Zoltan Fodor and Christian Hoelbling.
\newblock \href{https://journals.aps.org/rmp/abstract/10.1103/RevModPhys.84.449}{Light hadron masses from lattice QCD}.
\newblock {\em Reviews of Modern Physics}, 84(2):449, 2012.

\bibitem{grable2022theremal}
Seth Grable.
\newblock \href{https://link.springer.com/article/10.1007/JHEP10(2022)133}{Interacting CFTs for all couplings: thermal versus entanglement entropy at large N.}
\newblock {\em J. High Energ. Phys.}, 133, 2022.

\bibitem{grable2023elements}
Seth Grable and Paul Romatschke.
\newblock \href{https://arxiv.org/pdf/2310.12203}{Elements of Confinement for QCD with Twelve Massless Quarks}.
\newblock {\em arXiv preprint arXiv:2310.12203}, 2023.

\bibitem{grable2023fully}
Seth Grable and Max Weiner.
\newblock \href{https://link.springer.com/article/10.1007/JHEP09(2023)017}{A fully solvable model of fermionic interaction in 3+ 1d}.
\newblock {\em Journal of High Energy Physics}, 2023(9):1--14, 2023.

\bibitem{gross1973ultraviolet}
David~J Gross and Frank Wilczek.
\newblock \href{https://journals.aps.org/prl/abstract/10.1103/PhysRevLett.30.1343}{Ultraviolet behavior of non-abelian gauge theories}.
\newblock {\em Physical Review Letters}, 30(26):1343, 1973.

\bibitem{heisenberg2006consequences}
W~Heisenberg and H~Euler.
\newblock \href{https://arxiv.org/pdf/physics/0605038}{Consequences of dirac theory of the positron}.
\newblock {\em arXiv preprint physics/0605038}, 2006.

\bibitem{jack1982two}
I~Jack and H~Osborn.
\newblock \href{https://www.sciencedirect.com/science/article/pii/0550321382902127?fr=RR-1&ref=cra_js_challenge}{Two-loop background field calculations for arbitrary background fields}.
\newblock {\em Nuclear Physics B}, 207(3):474--504, 1982.

\bibitem{laermann2003lattice}
Edwin Laermann and Owe Philipsen.
\newblock \href{https://www.annualreviews.org/content/journals/10.1146/annurev.nucl.53.041002.110609}{Lattice QCD at finite temperature}.
\newblock {\em Annual Review of Nuclear and Particle Science}, 53(1):163--198, 2003.

\bibitem{laine2016basics}
Mikko Laine and Aleksi Vuorinen.
\newblock \href{https://link.springer.com/content/pdf/10.1007/978-3-319-31933-9.pdf}{Basics of thermal field theory}.
\newblock {\em Lect. Notes Phys}, 925(1):1701--01554, 2016.

\bibitem{leutwyler1981constant}
H~Leutwyler.
\newblock \href{https://www.sciencedirect.com/science/article/pii/0550321381902522}{Constant gauge fields and their quantum fluctuations}.
\newblock {\em Nuclear Physics B}, 179(1):129--170, 1981.

\bibitem{mathieu2009physics}
Vincent Mathieu, Nikolai Kochelev, and Vicente Vento.
\newblock \href{https://www.worldscientific.com/doi/abs/10.1142/S0218301309012124?casa_token=24LECYea28kAAAAA%3Alg5MxGYu4ylEucbPkU3rG05AgJywSbH-lT_9BZ5EcBqVjzUq_AGdQEKVsdo98dm4Fz86cGjzLDY}{The physics of glueballs}.
\newblock {\em International Journal of Modern Physics E}, 18(01):1--49, 2009.

\bibitem{morningstar1999glueball}
Colin~J Morningstar and Mike Peardon.
\newblock \href{https://journals.aps.org/prd/abstract/10.1103/PhysRevD.60.034509}{Glueball spectrum from an anisotropic lattice study}.
\newblock {\em Physical Review D}, 60(3):034509, 1999.

\bibitem{moshe2003quantum}
Moshe Moshe and Jean Zinn-Justin.
\newblock \href{https://www.sciencedirect.com/science/article/pii/S0370157303002631?casa_token=oouDxWyPCacAAAAA:9qdvt9nWLYZg0yLXUlwG-gBSY0zp1FfENmCT2nYEBGElFRQzHE6aDY3IW_dGh9r50Cy5ilAJgw}{Quantum field theory in the large N limit: A Review}.
\newblock {\em Physics Reports}, 385(3-6):69--228, 2003.

\bibitem{nielsen1979quantum}
Holger~Bech Nielsen and P~Olesen.
\newblock \href{https://www.sciencedirect.com/science/article/pii/0550321379900658}{A quantum liquid model for the QCD vacuum: gauge and rotational invariance of domained and quantized homogeneous color fields}.
\newblock {\em Nuclear Physics B}, 160(2):380--396, 1979.

\bibitem{nielsen1978unstable}
N\_K Nielsen and P~Olesen.
\newblock \href{https://www.sciencedirect.com/science/article/pii/0550321378903772}{An unstable Yang-Mills field mode}.
\newblock {\em Nuclear Physics B}, 144(2-3):376--396, 1978.

\bibitem{ochs2013status}
Wolfgang Ochs.
\newblock \href{https://iopscience.iop.org/article/10.1088/0954-3899/40/4/043001/meta?casa_token=1aXZiUAxAUYAAAAA:SLSiCWcTz3Mv5Z-Ag6nLifwBBjqaPHNzhsL1sQu_-iJVXliBtQ8ic6Z9qS7T3qK5HnVGCV186HPWSNDnnSiwMiBY}{The status of glueballs}.
\newblock {\em Journal of Physics G: Nuclear and Particle Physics}, 40(4):043001, 2013.

\bibitem{peskin2018introduction}
Michael~E Peskin.
\newblock {\em \href{https://www.taylorfrancis.com/books/mono/10.1201/9780429503559/introduction-quantum-field-theory-michael-peskin}{An introduction to quantum field theory}}.
\newblock CRC press, 2018.

\bibitem{Romatschke2022}
Paul Romatschke.
\newblock \href{https://arxiv.org/abs/2211.15683}{A solvable quantum field theory with asymptotic freedom in 3+1 dimensions}.
\newblock {\em arXiv:2211.15683 [hep-th]}.

\bibitem{romatschke2019finite}
Paul Romatschke.
\newblock \href{https://journals.aps.org/prl/abstract/10.1103/PhysRevLett.122.231603}{Finite-temperature conformal field theory results for all couplings: O (N) model in 2+ 1 dimensions}.
\newblock {\em Physical Review Letters}, 122(23):231603, 2019.

\bibitem{romatschke2019simple}
Paul Romatschke.
\newblock \href{https://link.springer.com/article/10.1007/JHEP03(2019)149}{Simple non-perturbative resummation schemes beyond mean-field: case study for scalar $\phi$4 theory in 1+ 1 dimensions}.
\newblock {\em Journal of High Energy Physics}, 2019(3):1--16, 2019.

\bibitem{romatschke2023loophole}
Paul Romatschke.
\newblock \href{https://arxiv.org/abs/2310.18414}{A loophole in the proofs of asymptotic freedom and quantum triviality}.
\newblock {\em arXiv preprint arXiv:2310.18414}, 2023.

\bibitem{romatschke2023if}
Paul Romatschke.
\newblock \href{https://www.sciencedirect.com/science/article/pii/S0370269323006044}{What if $\phi$4 theory in 4 dimensions is non-trivial in the continuum?}
\newblock {\em Physics Letters B}, 847:138270, 2023.

\bibitem{romatschke2024life}
Paul Romatschke.
\newblock \href{https://www.mdpi.com/2673-9909/4/1/3}{Life at the Landau pole}.
\newblock {\em AppliedMath}, 4(1):55--69, 2024.

\bibitem{romatschke2024mass}
Paul Romatschke, Chun-Wei Su, and Ryan Weller.
\newblock \href{https://arxiv.org/abs/2405.00088}{Mass from Nothing}.
\newblock {\em arXiv preprint arXiv:2405.00088}, 2024.

\bibitem{savvidy2020heisenberg}
George Savvidy.
\newblock \href{https://link.springer.com/article/10.1140/epjc/s10052-020-7711-6}{From Heisenberg--Euler Lagrangian to the discovery of chromomagnetic gluon condensation}.
\newblock {\em The European Physical Journal C}, 80(2):165, 2020.

\bibitem{savvidy2023stability}
George Savvidy.
\newblock \href{https://www.sciencedirect.com/science/article/pii/S0370269323004161}{On the stability of Yang-Mills vacuum}.
\newblock {\em Physics Letters B}, 844:138082, 2023.

\bibitem{savvidy2024large}
George Savvidy.
\newblock How large is the space of covariantly constant gauge fields.
\newblock {\em Nuclear Physics B}, 1004:116561, 2024.

\bibitem{savvidy1977infrared}
GK~Savvidy.
\newblock \href{https://www.sciencedirect.com/science/article/pii/0370269377907596}{Infrared instability of the vacuum state of gauge theories and asymptotic freedom}.
\newblock {\em Physics letters B}, 71(1):133--134, 1977.

\bibitem{schwinger1951gauge}
Julian Schwinger.
\newblock \href{https://journals.aps.org/pr/abstract/10.1103/PhysRev.82.664}{On gauge invariance and vacuum polarization}.
\newblock {\em Physical Review}, 82(5):664, 1951.

\bibitem{soley2023experimentally}
Micheline~B Soley, Carl~M Bender, and A~Douglas Stone.
\newblock \href{https://journals.aps.org/prl/abstract/10.1103/PhysRevLett.130.250404}{Experimentally realizable pt phase transitions in reflectionless quantum scattering}.
\newblock {\em Physical Review Letters}, 130(25):250404, 2023.

\bibitem{t1976computation}
Gerard t~Hooft.
\newblock \href{https://journals.aps.org/prd/abstract/10.1103/PhysRevD.14.3432}{Computation of the quantum effects due to a four-dimensional pseudoparticle}.
\newblock {\em Physical Review D}, 14(12):3432--3450, 1976.

\bibitem{t1973algorithm}
Gerardus t~Hooft.
\newblock \href{https://www.sciencedirect.com/science/article/pii/0550321373902630?via%3Dihub}{An algorithm for the poles at dimension four in the dimensional regularization procedure}.
\newblock {\em Nucl. Phys. B}, 62(CERN-TH-1692):444--60, 1973.

\bibitem{weinberg1995quantum}
Steven Weinberg.
\newblock {\em \href{https://books.google.com/books?hl=en&lr=&id=48xXMF1oHxkC&oi=fnd&pg=PR17&dq=theory+of+quantum+fields+volume+II+wienberg&ots=nkiuZl-c6z&sig=higYWEq2amEK6qXxbIx5NTzohHs#v=onepage&q=theory%20of%20quantum%20fields%20volume%20II%20wienberg&f=false}{The quantum theory of fields}}, volume~2.
\newblock Cambridge university press, 1995.

\bibitem{wilson1974confinement}
Kenneth~G Wilson.
\newblock \href{https://journals.aps.org/prd/abstract/10.1103/PhysRevD.10.2445}{Confinement of quarks}.
\newblock {\em Physical review D}, 10(8):2445, 1974.

\bibitem{yildiz1980vacuum}
Asim Yildiz and Paul~H Cox.
\newblock \href{https://journals.aps.org/prd/abstract/10.1103/PhysRevD.21.1095}{Vacuum behavior in quantum chromodynamics}.
\newblock {\em Physical Review D}, 21(4):1095, 1980.

\end{thebibliography}
\end{spacing}
\end{document}